\documentclass[aps,prd,superscriptaddress,long,notitlepage,balancelastpage,nofootinbib,floatfix,twocolumn]{revtex4-1}
\pdfoutput=1
\usepackage{amsmath,mathtools,amssymb,amsthm,amsxtra,overpic,bbm,epsfig,subfigure,url,bm}
\usepackage{hyperref}
\usepackage{mathrsfs}
\usepackage{color,xcolor}
\usepackage{comment}
\usepackage{float}
\usepackage{enumitem}
\usepackage{slashed}
\usepackage{multirow}
\usepackage{appendix}
\DeclareUnicodeCharacter{03BC}{\ensuremath{\mu}}
\DeclareUnicodeCharacter{2032}{\ensuremath{'}}
\definecolor{nicered}{rgb}{0.5,0.,0.}
\definecolor{nicegreen}{rgb}{0.,0.5,0.}
\definecolor{niceblue}{rgb}{0.,0.,0.5}
\hypersetup{
	colorlinks=true,
	linkcolor=black,
	filecolor=nicegreen,      
	urlcolor=niceblue,
	citecolor=nicered,
}

\makeatletter
\newcommand*{\balancecolsandclearpage}{%
	\close@column@grid
	\cleardoublepage
	\twocolumngrid
}
\makeatother
\begin{document}

\title{\vspace{1cm} \Large 
 Muonic Force and Neutrino Non-Standard \\ Interactions  at Muon Colliders
}

\author{\bf Sudip Jana}
\email[E-mail:]{sudip.jana@mpi-hd.mpg.de}
\affiliation{Max-Planck-Institut f{\"u}r Kernphysik, Saupfercheckweg 1, 69117 Heidelberg, Germany}

\author{\bf Sophie Klett}
\email[E-mail:]{sophie.klett@mpi-hd.mpg.de}
\affiliation{Max-Planck-Institut f{\"u}r Kernphysik, Saupfercheckweg 1, 69117 Heidelberg, Germany}

\begin{abstract}
The discovery of neutrino oscillations implies that neutrinos are massive and mixed, necessitating an extension of the Standard Model, which may require the introduction of non-standard neutrino interactions (NSI). We investigate the potential of a high-energy muon collider to probe such NSIs with muons, specifically focusing on muonic forces. By analyzing the monophoton signal from the process $\mu^+ \mu^- \rightarrow \nu \overline{\nu } \gamma$, we explore four-fermion contact interactions involving two muons and two neutrinos. Moreover, we examine minimal models that generate scalar and vector-mediated NSIs and study their phenomenology. Projected sensitivities for the strength of NSIs, $\epsilon_{\alpha \beta}^{\mu \mu}$, are presented at a 95$\%$ confidence level for a center-of-mass energy of 3 TeV and integrated luminosities of $\mathcal{L} = 1$ and $10~\mathrm{ab}^{-1}$, showcasing the complementarity between a muon collider and other experimental probes.
\noindent 
\end{abstract}
\maketitle
\textbf{\emph{Introduction}.--} 
In the 1930s, Anderson and Neddermeyer made a groundbreaking observation of charged long-lived particles that were ``less massive than protons but more penetrating than electrons," \cite{Anderson:1936zz, Neddermeyer:1937md}   leading to the momentous discovery of the muon and marking a pivotal moment in modern particle physics. Despite significant strides in understanding particle physics and the remarkable success of the Standard Model (SM), there remain compelling unresolved issues involving muon sectors. Historically, the muon has played a vital role in our understanding; for example, the most precise determination of the Fermi constant $G_F$ relies on muon decay rate measurements. Recently, a series of intriguing muon measurements has sparked renewed optimism, with tantalizing hints of physics beyond the SM, particularly revolving around the anomalous magnetic moment of the muon \cite{Abi:2021gix, Bennett:2006fi} and rare $B$ meson decay observables \cite{Aaij:2021vac, Aaij:2017vbb, Aaij:2019wad}. These observables could potentially point towards the existence of a new muonic force. 

The discovery of neutrino oscillations provides compelling evidence for new physics beyond the Standard Model, as it necessitates the existence of tiny neutrino masses and mixings. Constructing neutrino mass models involves considering new interactions of neutrinos, referred to as non-standard neutrino interactions (NSI). A recent study includes the classification of various models of neutrino mass generation into type-I and type-II based on the NSIs \cite{Babu:2019mfe}. Most studies \cite{Proceedings:2019qno} have focused on the interaction of neutrinos with the first generation of charged leptons and quarks, namely electrons, up, and down quarks, which has the most significant impact on neutrino propagation through matter \cite{Wolfenstein:1977ue} and potentially observable effects in neutrino oscillation experiments. There have been numerous studies to investigate the NSIs in neutrino oscillation and scattering experiments \cite{Proceedings:2019qno, Coloma:2023ixt, Esteban:2018ppq}, neutrino telescopes \cite{Babu:2019vff, Babu:2022fje, Huang:2021mki} and collider experiments \cite{Babu:2020nna, Liu:2020emq, Choudhury:2018xsm, Friedland:2011za, BuarqueFranzosi:2015qil, Barranco:2007ej, Berezhiani:2001rs}. However, this article uniquely focuses on neutrino interactions with the second generation of charged leptons (muons) and explores their implications.
The exploration of four-fermion contact interactions featuring a pair of muons and two neutrinos are driven by the intriguing potential that the novel physics underlying neutrino mass mechanisms might also  offer tantalizing hints of novel phenomena involving muons, thereby necessitating the presence of these interactions. These muonic NSI can  also significantly impact astrophysical environments \cite{Croon:2020lrf, Caputo:2021rux, Manzari:2023gkt}, particularly in proto-neutron stars formed during core-collapse supernovae, where there is a  sizable population of muons.

As the design for a high-energy muon collider enters a new phase \cite{MuonCollider:2022nsa, Accettura:2023ked, MuonCollider:2022glg, Aime:2022flm, MuonCollider:2022xlm}, it becomes crucial to consider the potential for discovering new physics. The muon collider holds great promise for exploring the complementarity between energy and precision, making it an effective single collider that combines the advantages of $ee$ and $pp$ machines. Unlike protons, muons are point-like particles, allowing the full nominal center-of-mass collision energy $E_{cm}$ to be harnessed for producing high-energy reactions, probing length scales as short as 1$/E_{cm}$. Recently, there has been a rising interest in exploring new physics at muon colliders \cite{Han:2020uak, Bandyopadhyay:2020otm,Capdevilla:2021fmj,Bottaro:2021snn,Bottaro:2022one,Kalinowski:2022fot, Homiller:2022iax, Chakrabarty:2014pja,Costantini:2020stv,Buttazzo:2020uzc,Han:2020pif,Chiesa:2020awd,Han:2021udl,Forslund:2022xjq,Belfkir:2023lot,Chakraborty:2022pcc,Li:2023tbx,Kwok:2023dck,Capdevilla:2020qel,Buttazzo:2020ibd,Capdevilla:2021kcf,Dermisek:2021mhi,Capdevilla:2021rwo, Huang:2021biu,Altmannshofer:2022xri,  Asadi:2021gah, Azatov:2022itm, Altmannshofer:2023uci,Sun:2023cuf, Chowdhury:2023imd, Maharathy:2023dtp} (see \cite{MuonCollider:2022xlm} for an exhaustive list of references).
Here, we investigate neutrino non-standard interactions involving muons, and the muon collider presents an excellent platform for exploring these interactions. For illustration, we explore minimal models generating such scalar and vector-mediated muonic NSIs: (a) a new muon-philic scalar model, motivated as one explanation for the observed discrepancy between the measured and predicted muon magnetic moment, while also addressing the large neutrino self-interaction solution to the Hubble tension; (b) the popular Zee model, where muonic NSIs arise from the exchange of charged scalars; and (c) the $L_\mu - L_\tau$ model, generating muonic NSIs through the exchange of a new neutral gauge boson $Z'$.

The article is organized as follows: Section II defines EFT and lists muonic NSI-related operators. In Section III, a simplified scenario is analyzed for the muon collider's sensitivity to muonic NSIs. Section IV showcases illustrative ultraviolet (UV) completions and their complementarity with other experimental probes. Finally, Section V provides the conclusions.
\vspace{0.1 in}

\textbf{\emph{From EFTs to Simplified Models}.--} 

	\begin{figure}[ht!]
		\begin{center}
			\includegraphics[width=0.48\textwidth]{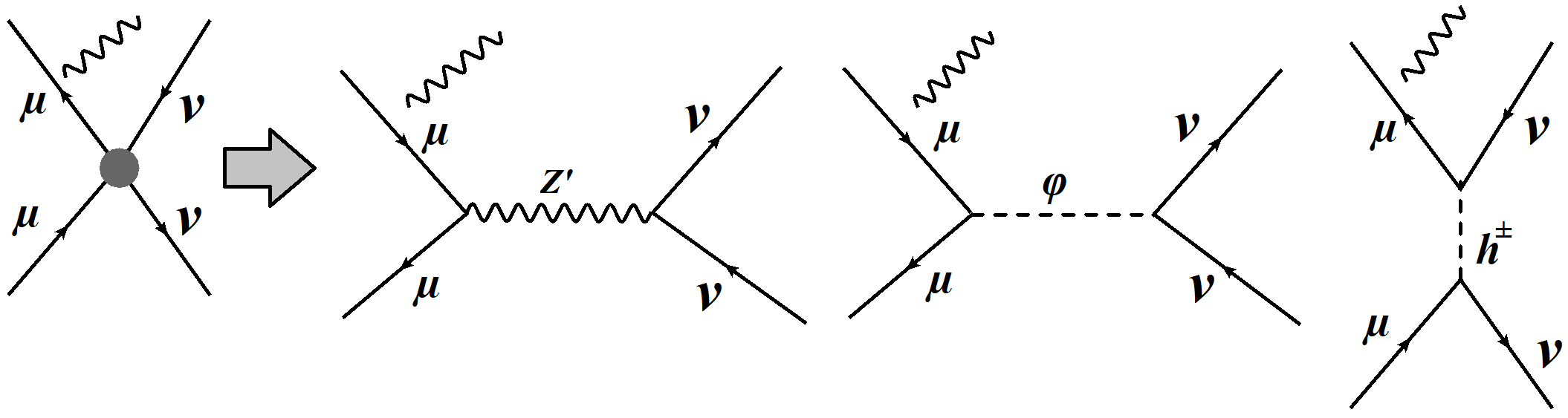}
		\end{center}
		\vspace{-0.2cm}
		\caption{Illustrative Feynman diagrams depicting the monophoton signal at a muon collider in the $\mu^+ \mu^- \rightarrow \nu \overline{\nu } \gamma$ process, arising from NSI $\epsilon_{\alpha \beta}^{\mu \mu}$.}
		\label{fig:schematic}
	\end{figure}
Neutrino non-standard interactions with two muons, in their most general form, can be parameterized by a set of dimension six effective operators, including scalar, pseudoscalar, vector, axial-vector and tensor four-fermion interactions. The corresponding Lagrangian, normalized to the Fermi constant $G_F$, can be expressed as:
\begin{equation}
\label{Eq.: NSI operator general}
\begin{split}
    \mathcal{L}_{\mu \mu  \nu \nu}& = -2 \sqrt{2} G_F \sum_{\alpha, \beta} \sum_{i=1}^{10} {\epsilon_{\alpha \beta}^{\mu \mu }}^{(i)} \left(\overline{\nu}_\alpha \mathcal{O}_i \nu_{\beta} \right) \left(\overline{\mu} \mathcal{O}'_i \mu\right)
    ~,
\end{split}
\end{equation}
where $\alpha, \beta \in \{e,~\mu,~\tau\}$ denote the neutrino flavors. The operators $\mathcal{O},~ \mathcal{O}' $ and effective couplings ${\epsilon_{\alpha \beta}^{\mu \mu }}^{(i)}$ are listed in Table \ref{Tab.: Effective Operators} where we use the usual convention for chiral projection operators $P_{L/R} \equiv \left(1\mp\gamma^5\right)/2$ and $\sigma^{\mu \nu}\equiv \frac{i}{2}\left[\gamma^\mu, \gamma^\nu \right]$. 
Considering the SM, the only non-vanishing contribution originates from the neutral current and charged current interaction  giving rise to the flavor diagonal coefficients
\begin{equation}
\begin{split}
    {\epsilon_{\alpha \beta}^{\mu \mu }}^{(V,LL)} &=\delta_{\alpha \mu} \delta_{\beta \mu} +\left(- \dfrac{1}{2} + \sin^2{\theta_w} \right) \delta_{\alpha \beta} \indent (\mathrm{SM}) ~,\\
    {\epsilon_{\alpha \beta}^{\mu \mu }}^{(V,LR)} &= \sin^2{\theta_w}\delta_{\alpha \beta} \indent \indent\indent\indent\indent\indent\indent  \indent\indent ~~ ~~ (\mathrm{SM})  ~,
\end{split}
\end{equation}
with the Weinberg angle $\theta_w$.
Manifestly, the interactions of the $Z$-boson are common to all neutrino flavors, whereas the contribution from the $W$-boson is restricted to $\nu_\mu$ only.
\begin{table}[t]
\begin{tabular}{|c|c|c|c|}
\hline
$\boldsymbol{i}$ & $\boldsymbol{{\epsilon_{\alpha \beta}^{\mu \mu }}^{(i)}}$ &$\boldsymbol{\mathcal{O}} $&$\boldsymbol{\mathcal{O}'} $\\
\hline
$1$  & ${\epsilon_{\alpha \beta}^{\mu \mu }}^{(V,LL)} $   & $\gamma_\mu P_L$      &  $\gamma^\mu P_L$  \\
$2$  & ${\epsilon_{\alpha \beta}^{\mu \mu }}^{(V, RL)} $   & $\gamma_\mu P_R$      &  $\gamma^\mu P_L$  \\
$3$  & ${\epsilon_{\alpha \beta}^{\mu \mu }}^{(V, LR)} $   & $\gamma_\mu P_L$      &  $\gamma^\mu P_R$  \\
$4$  & ${\epsilon_{\alpha \beta}^{\mu \mu }}^{(V, RR)} $   & $\gamma_\mu P_R$      &  $\gamma^\mu P_R$  \\
$5$  & ${\epsilon_{\alpha \beta}^{\mu \mu }}^{(S, LL)} $   & $ P_L$      &  $ P_L$  \\
$6$  & ${\epsilon_{\alpha \beta}^{\mu \mu }}^{(S, RL)} $   & $ P_R$      &  $P_L$  \\
$7$  & ${\epsilon_{\alpha \beta}^{\mu \mu }}^{(S, LR)} $   & $ P_L$      &  $ P_R$  \\
$8$  & ${\epsilon_{\alpha \beta}^{\mu \mu }}^{(S, RR)} $   & $ P_R$      &  $ P_R$  \\
$9$  & ${\epsilon_{\alpha \beta}^{\mu \mu }}^{(T, LL)} $   & $\sigma_{\mu \nu} P_L$      &  $ \sigma^{\mu \nu} P_L$  \\
$10$  & ${\epsilon_{\alpha \beta}^{\mu \mu }}^{(T, LL)} $   & $\sigma_{\mu \nu} P_R$      &  $ \sigma^{\mu \nu} P_R$  \\
\hline
\end{tabular}
\caption{List of effective operators and couplings that contribute to the four-fermion interaction $\mu \mu \nu \nu$ up to dimension six.}
\label{Tab.: Effective Operators}
\end{table}

While the EFT approach can be suitable to describe oscillation experiments at low energy scales, the momentum transfer at colliders can be sizable, and the applicability of EFT in these regimes is questionable. For our collider study, we follow an elaborate approach where we include explicit force mediators in our analysis. In detail, we discuss three types of BSM scenarios that can lead to NSI with muons, namely neutral gauge bosons and neutral or  charged scalars, as depicted in Fig.~\ref{fig:schematic}. To begin with, we examine a simplified $Z'$  model and stay agnostic about possible UV completions for now. More concrete examples will be discussed in the following sections.

We describe the $Z'$ mediated NSIs by the Lagrangian
\begin{equation}
\label{Eq.: Z' simplified model lagrangian}
\mathcal{L}_{\mathrm{NSI}}^{\mathrm{Simp}, Z'} = \sum_{Y=L,R} \left[(g_\nu)_{\alpha \beta} \overline{\nu}_\alpha \gamma^\mu P_L \nu_\beta +  g^{Y}_{\mu} \overline{\mu} \gamma^\mu P_Y \mu\right] Z'_\mu ~,
\end{equation}
where $g_\nu$ and $g_\mu^Y$ denote the $Z'$ couplings to neutrinos and muons respectively. Integrating out the $Z'$ boson yields a contribution to the operators $i=1,3$, and the corresponding effective couplings can be identified as 
\begin{equation}
    \begin{split}
        {\epsilon^{\mu \mu}_{\alpha \beta}}^{(V,LY)} &= \frac{(g_\nu)_{\alpha \beta} g_\mu^Y}{2 \sqrt{2} G_F M_{Z'}^2} ~, \indent Y \in \{L,R\} ~,
    \end{split}
\end{equation}
where we treat the $Z'$ mass $M_{Z'}$ as a free parameter. In what follows, we describe how a muon collider can probe these parameters. 

\begin{figure}[htb!]
		\centering
		\includegraphics[width=0.4\textwidth]{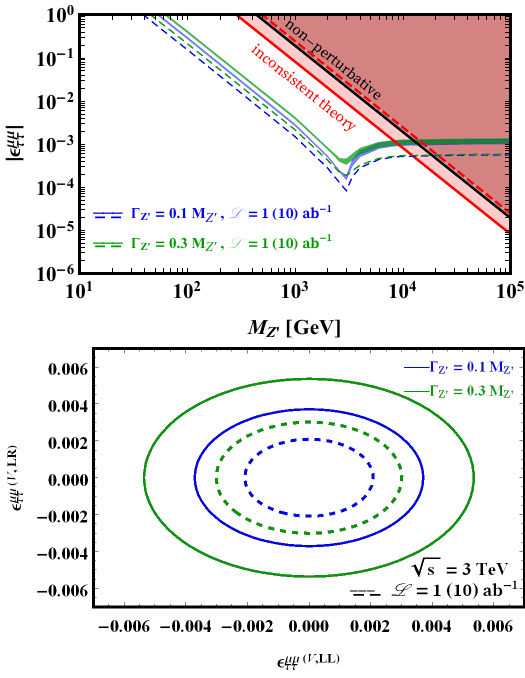}
	\caption{Sensitivity to NSI parameters for a $\sqrt{s} = 3$ TeV muon collider with integrated luminosity $\mathcal{L}= 1,~10~ \mathrm{ab}^{-1}$ at $95\%$ CL. The upper plot shows the sensitivity contours for $|\epsilon_{\tau\tau}^{\mu\mu}|$ with a vector-like coupling to muons and a  mediator decay width $\Gamma_{Z'} = 0.1 (0.3) M_{Z'}$ in blue (green) color. In the lower plot we display the degeneracy in the $\left(\epsilon^{\mu\mu (V,LL)}_{\tau\tau}, \epsilon^{\mu\mu (V,LR)}_{\tau\tau}\right)$ plane for a mediator mass $M_{Z'} = 1 $ TeV.}
	\label{Fig:Sensitivity simplified models}
\end{figure}

\vspace{0.05in}
\textbf{\emph{Probing NSI from Monophoton Signal}.--}
By scrutinizing  mono-X events (monophoton, mono-jet, mono-$Z$, mono-$W$) with notable missing transverse energy, akin to dark matter searches, one can probe indications of neutrino NSIs in collider experiments. Here we investigate NSI by analyzing the monophoton signal, $\mu^+ \mu^- \rightarrow   \gamma +\slashed{E}_T $, at the muon collider. For our analysis, we use {\tt FeynRules} package \cite{Alloul:2013bka} for model implementation and simulate signal and background events for the process $\mu^+ \mu^- \rightarrow   \nu \overline{\nu}\gamma $ with the Monte Carlo event generator {\tt MadGraph5aMC@NLO} \cite{Alwall:2014hca, Alwall:2011uj} interfaced to {\tt Pythia8} \cite{Bierlich:2022pfr} for the parton showering and hadronization. We perform  detector simulation with the {\tt Delphes3} package \cite{deFavereau:2013fsa}. For the event selection, we require the following criteria for  photon  transverse momentum and  pseudorapidity:
\begin{equation}
   p_\gamma^T > 10 ~\mathrm{GeV}~, \indent|\eta_\gamma |< 2.44
\end{equation}
with the last requirement being equivalent to a detector acceptance $10^{\circ} < \theta_\gamma < 170^{\circ}$. 
Generally, there can be interference between SM contributions and NSIs for flavor diagonal couplings. However, we find the interference effect negligible for the cases considered here, as can be seen from Fig.~\ref{Fig:Interference cross check} in the supplemental material. Consequently, the signal cross section scales as $\propto \epsilon^2$, which renders the parameter $\epsilon$ to a suitable quantity to describe the sensitivity of a muon collider to NSIs \cite{Berezhiani:2001rs}.
We  partition our sample events into nine separate signal regions based on $\slashed{E}_T$ 
 acceptance criteria:
\begin{equation}
    \slashed{E}_T > 10,~20,~30,~40,~50,~60,~70,~80, 100~\mathrm{GeV}
\end{equation}
and  obtain the NSI sensitivity from the signal region, which maximizes the statistical significance  
\begin{equation}
    \mathcal{S} \equiv \dfrac{N_S}{\sqrt{ N_B + N_S+ (\delta \sigma_B N_B)^2}}
\end{equation}
with $N_S$ ($N_B$) the number of signal (background) events and a systematic uncertainty $\delta \sigma_B$ on the background. In Fig.~\ref{Fig:Sensitivity simplified models}, we have shown the projected sensitivity of a muon collider to NSIs at $95\%$ CL for a   centre-of-mass energy $\sqrt{s}= 3$ TeV   with two different integrated luminosities $\mathcal{L} = 1,~10~\mathrm{ab}^{-1} $  and for the decay widths $\Gamma_{Z'}/M_{Z'} = 0.1$ and $ 0.3$. For simplicity, we consider the coupling to tau neutrinos to be the only non-vanishing one, i.e. $ (g_\nu)_{\alpha \beta}\equiv g_\nu \delta_{\alpha \tau} \delta_{\beta \tau} $. The upper plot shows the sensitivity to the NSI parameter $ |\epsilon_{\tau\tau}^{\mu\mu(V,LL)} |$ in the simplified $Z'$ model assuming pure vector-like muon coupling $g_\mu^L = g_\mu^R$. Hereinafter, for  brevity, we drop the superscripts in brackets and write $|\epsilon_{\tau\tau}^{\mu\mu}| $ whenever it is clear from the context. For the low luminosity case, we present the impact of the systematic uncertainty $\delta \sigma_B = 0.0\% -0.1\%$ as a band, while for the high luminosity case we only display $\delta \sigma_B = 0.0\%$ \cite{Han:2020uak}. The highest sensitivity is reached when mediators are produced resonantly at $M_{Z'} \sim 3$ TeV. For the case $\mathcal{L}= 1~ \mathrm{ab}^{-1}$ and $\Gamma_{Z'}=0.1 M_{Z'}$ this would allow to  constrain $|\epsilon_{\tau\tau}^{\mu\mu} | \lesssim 1.5 \times 10^{-4}$. For large mediator masses the NSI bound is approximately constant, indicating the beginning of the EFT regime.
Since the total decay width of the $Z'$ is limited by its partial decay widths $\Gamma_{Z'} \geq \Gamma_{Z' \rightarrow \overline{\mu} \mu} + \Gamma_{Z' \rightarrow \overline{\nu} \nu}  $
we find the following constraint for a decay to a single neutrino flavor
\begin{equation}
\begin{split}
    \Gamma_{Z'} &\geq \dfrac{M_{Z'}}{24 \pi }\left[\left(g_\nu\right)^2 + \left(g_\mu^L\right)^2 + \left(g^R_\mu \right)^2 \right] 
    \geq \dfrac{M_{Z'}}{24 \pi } 2 \sqrt{2} g_\nu g_\mu^L
    ~,
\end{split}
\end{equation}
and we neglect phase space factors of the order $\mathcal{O}(m_f^2/M^2_{Z'})$ where $f=\nu, ~\mu$ (see for example \cite{Babu:2020nna} for full expression). By fixing the ratio $\Gamma_{Z'}/M_{Z'}$ we immediately obtain  a consistency constrain for the NSI parameter  that is given by
\begin{equation}
    |\epsilon_{\tau\tau}^{\mu\mu} | \leq \dfrac{3 \pi}{G_F M_{Z'}^2} \dfrac{\Gamma_{Z'}}{M_{Z'}}~. 
\end{equation}
For the case $\Gamma_{Z'}= 0.1 M_{Z'}$  the consistency requirement (solid red line) is more stringent than the perturbativity limit where $g_\nu g_\mu^L = 2 \pi$ (solid black line). Note that, broadening the resonance $\Gamma_{Z'}= 0.3 M_{Z'}$ alleviates the consistency constraint (dashed red line).
In the lower plot of Fig.~\ref{Fig:Sensitivity simplified models}, we  show the sensitivity on the parameters $\epsilon^{\mu\mu (V,LL)}_{\tau\tau}$ vs. $ \epsilon^{\mu\mu (V,LR)}_{\tau\tau}$ for  a gauge boson mass $M_{Z'}=1$ TeV with arbitrary couplings $g_\mu^L, ~g_\mu^R$. Notably, this result demonstrates that constraints on a $Z'$ model with purely vector-like couplings to muons $(g_\mu^L=g_\mu^R)$  can be as strong as constraints on a model with only axial-vector couplings $(g_\mu^L=-g_\mu^R)$.

\vspace{0.05 in}
\textbf{\emph{UV complete Models}.--}
When constructing UV complete models for observable NSIs, the coexistence of charged leptons in the same SU(2)$_L$ multiplet as neutrinos add complexity, leading to potential charged lepton flavor violations and other extensive phenomenological consequences. As illustrative examples, we highlight three specific UV complete models resulting in non-standard interactions of neutrinos with muons (see categorization in Fig.~\ref{fig:schematic}).
\begin{figure*}[htb!]
		\centering
		\includegraphics[width=\textwidth]{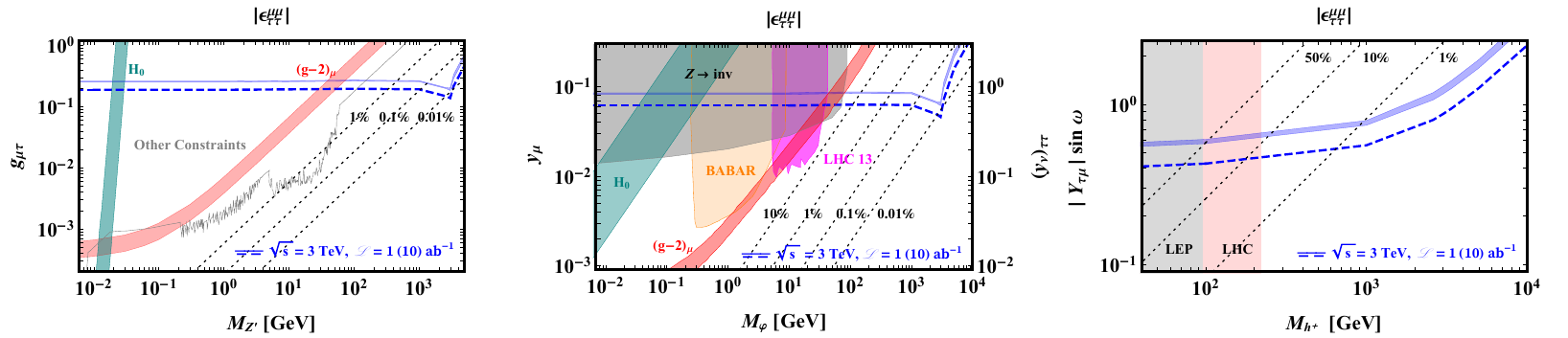}
	\caption{Sensitivity reach of a $\sqrt{s}=3$ TeV muon collider with integrated luminosity $\mathcal{L}= 1,~10 ~\mathrm{ab}^{-1}$  in the coupling  vs. mediator mass plane (blue solid and dashed line) together with existing experimental bounds. Dotted lines represent isocontours  for the NSI parameter $|\epsilon_{\tau \tau}^{\mu\mu}|$. Left: gauged $L_\mu - L_\tau$ model, 
    middle: neutral scalar extension of the SM, right: singly charged scalar extension. For more details, see the main text.
 }
	\label{Fig: Sensitivity UV models}
\end{figure*} 

First, we focus on the gauged $L_\mu - L_\tau$ model, which not only gauges an anomaly-free $U(1)$ symmetry of the SM \cite{He:1990pn, He:1991qd} but also can potentially address issues like the muon's anomalous magnetic moment and the Hubble tension \cite{Baek:2001kca, Ma:2001md, Amaral:2020tga, Escudero:2019gzq}. The relevant part of the Lagrangian can be expressed as:
\begin{equation}
\label{Eq.: Lmu Ltau model}
\mathcal{L}\supset 
 g_{\mu \tau}\left[ \overline{L}_\mu \gamma^\mu  L_\mu +  \overline{\mu}_R \gamma^\mu  \mu_R
 -\overline{L}_\tau \gamma^\mu  L_\tau
 -  \overline{\tau}_R \gamma^\mu  \tau_R
 \right] Z'_\mu ~,
\end{equation}
where $g_{\mu \tau}$ is the new gauge coupling and $ L_\mu \equiv \left(\nu_{\mu L}, \mu_L \right)^T$,  $ L_\tau \equiv \left(\nu_{\tau L}, \tau_L \right)^T$ denote the left-handed weak lepton doublets. Following the analysis technique from the previous section,  Fig.~\ref{Fig: Sensitivity UV models}  presents the  $95 \%$ CL exclusion limit on the $g_{\mu\tau}$ vs. $M_{Z'}$ parameter space from the monophoton signal search at a $\sqrt{s} = 3$ TeV muon collider with integrated luminosities $\mathcal{L} = 1, ~10 ~ \mathrm{ab}^{-1}$. For comparison, we  display the parameter regions resolving $(g-2)_\mu$ at $2\sigma$ (red band) \cite{Amaral:2020tga} and the Hubble tension with $\Delta N_{\mathrm{eff}} \simeq 0.2-0.5$ (green band) \cite{Escudero:2019gzq}. A summary of current experimental constraints including CMS and BaBar four-muon searches \cite{CMS:2018yxg,BaBar:2016sci}, CCFR limit on neutrino trident production \cite{Altmannshofer:2014pba}, white Dwarf cooling \cite{Bauer:2018onh} and $N_{\mathrm{eff}}$ from  cosmological observations \cite{Escudero:2019gzq} is shown as a grey band. Contour lines of the NSI coefficient $\epsilon_{\tau \tau}^{\mu\mu} = g_{\mu \tau}^2/(2 \sqrt{2} G_F M_{Z'}^2)$ are displayed as black dotted lines. Clearly, the most sensitive region to test NSIs is at $M_{Z'}\sim 3$ TeV where it is possible to set a limit  $|\epsilon_{\tau \tau}^{\mu\mu}| \lesssim 1.1 \times 10^{-4} $ for a luminosity $\mathcal{L} = 1 ~\mathrm{ab}^{-1}$. Our result shows that the monophoton search at a muon collider can test interesting parts of the parameter space for the gauged $L_\mu-L_\tau$ model, however it is not competitive with current experimental bounds for $M_{Z'} \leq  100$ GeV. Yet, it can give some complementary insights in directly probing the gauge coupling  in a laboratory experiment. Note also that much stricter constraints for $M_{Z'} \geq  100$ GeV can arise  by considering processes like $\mu^+ \mu^- \rightarrow \overline{f} f \gamma$ where $f= \mu, \tau$  \cite{Huang:2021nkl}.

Now, we investigate a scalar extension of the SM, which is motivated as a potential explanation for the observed discrepancy between the measured and predicted muon magnetic moment while also addressing the large neutrino self-interaction solution to the Hubble tension.
\begin{equation}
\label{Eq.: Scalar Model}
    -\mathcal{L}_Y \supset \sum_{\alpha \beta}(y_\nu)_{\alpha \beta} \overline{\nu}_{\alpha}  P_L \nu_\beta  \varphi + y_\mu \overline{\mu} P_L \mu\varphi + h.c.~ ,
\end{equation}
where $(y_\nu)_{\alpha \beta}$ and $y_\mu$ represent Yukawa couplings to  a neutral scalar $\varphi$.
This type of scalar interaction with muons can naturally occur in two-Higgs-doublet models (2HDMs), offering at the same time the  possibility to explain the anomalous $(g-2)_\mu$ \cite{Jana:2020pxx}. Concerning the coupling to left-handed neutrinos, such type of interaction could, for instance, arise from the type-II seesaw  mechanism \cite{Magg:1980ut, Schechter:1980gr, Lazarides:1980nt, Mohapatra:1980yp}, where the neutral component of a weak triplet scalar mixes with the neutral component of the second Higgs doublet.
Even though the Yukawa couplings to muons and neutrinos can be arbitrary in general, we follow a simplified assumption where $\varphi$ only couples to the tau neutrino flavor and we set $(y_\nu)_{\tau \tau}/10=y_\mu$. Fig.~\ref{Fig: Sensitivity UV models} shows the result of our sensitivity study in the $(M_\varphi,y_\mu)$ plane, where isocontours of the NSI parameter $\epsilon_{\tau \tau}^{\mu\mu}\equiv (y_\nu)_{\tau\tau} y_\mu/(2 \sqrt{2} G_F M_\varphi^2)$ are shown as black dotted lines. The parameter space that can explain $(g-2)_\mu$ \cite{Jana:2020pxx} at $2\sigma$ and the Hubble tension \cite{Blinov:2019gcj} is highlighted with red and green color. For scalar masses $M_\varphi > 200$ MeV, BaBar limits the $\mu^+ \mu^- \varphi$ coupling by  $e^+ e^- \rightarrow \mu^+ \mu^- \varphi $ searches which we indicate as orange domain \cite{BaBar:2016sci,Batell:2017kty}. The parameter space that is excluded by CMS four-muon searches \cite{CMS:2018yxg,Batell:2017kty} and invisible $Z$ decay width \cite{Brdar:2020nbj} is shown as pink and grey shaded region. Note that the constraints from neutrino trident production could be relaxed for tau neutrinos. 
Our work demonstrates that the monophoton study at a muon collider can directly investigate the parameter region where a scalar with $M_\varphi \gtrsim 40$ GeV would explain the $(g-2)_\mu$ anomaly and can probe NSI strength as large as  $|\epsilon^{\mu\mu}_{\tau \tau}| \gtrsim 1.3 \times 10^{-4} $.

We now focus on NSI mediated by singly charged scalars which appear in many SM extensions. As a prototypical example, we analyse the Zee model \cite{Zee:1980ai}, which induces Majorana masses for neutrinos at the one-loop level.  The relevant part of the Lagrangian (see Supplemental Material for details) that will cause neutrino-muon interaction is:
\begin{equation}
-\mathcal{L}_Y \supset Y_{\alpha \beta}\left(h^{-} \sin \omega+H^{-} \cos \omega\right) \nu_\alpha \ell_\beta^c+\text { H.c. }
\end{equation}
where $\beta=\mu$. 
This leads to neutrino NSI with muons, which is given by 
\begin{equation}
 \epsilon_{\alpha \beta}^{\mu \mu }=\frac{Y_{\alpha \mu} Y_{\beta \mu}^{\star}}{4 \sqrt{2} G_F}\left(\frac{\sin ^2 \omega}{M_{h^{+}}^2}+\frac{\cos ^2 \omega}{M_{H^{+}}^2}\right)
\end{equation}

In the subsequent analysis, we assume simplified coupling of $h^+$ with tau neutrinos, setting $Y_{e\mu} = Y_{\mu\mu} = 0$ while keeping $Y_{\tau\mu} \neq 0$, and considering $M_{h^{+}} << M_{H^{+}}$. Fig.~\ref{Fig: Sensitivity UV models} shows NSI sensitivity using a 3 TeV muon collider with luminosity $\mathcal{L} = 1, 10~ \mathrm{ab}^{-1}$. It illustrates $95\%$ CL exclusion limits for $|Y_{\tau \mu}| \sin\omega$ vs. $M_{h^+}$, along with dashed lines indicating  $|\epsilon^{\mu \mu}_{\tau \tau}|$ contours induced by $h^+$. Notably, scalar masses under $95$ GeV are excluded by LEP searches \cite{Babu:2019mfe}, and current LHC smuon searches \cite{ATLAS:2022hbt} limit charged scalar masses up to 219 GeV. Unlike previous scenarios, here, NSI involves a $t$-channel process, lacking the resonance behavior near the center-of-mass energy. Our findings suggest that a 3 TeV muon collider could explore  Yukawa couplings down to $|Y_{\tau \mu}| \sin \omega \approx 0.45$ at $M_{h^+} \approx 220~\mathrm{GeV}$, allowing testing of NSI parameters $|\epsilon^{\mu\mu}_{\tau \tau}| \gtrsim 7\%$.

Subsequently, we judiciously fill in the remaining entries of the Yukawa matrices to explore their potential to generate both neutrino oscillation data and observable NSI with muons. We also ensure compatibility with experimental limits related to charged lepton flavor violation and other observables (see Supplementary Material for detailed discussion). For illustration, assuming real Yukawa couplings and normal mass ordering, a benchmark point that can reproduce observed neutrino oscillation data is given by:
\begin{equation}
Y_{\alpha \beta}=
\left(
\begin{array}{ccc}
 0. & 0. & -0.00378 \\
 6.747\times 10^{-6} & 0. & -3.219\times 10^{-6} \\
 0.000115 & 2.219 & -0.143 \\
\end{array}
\right)
\end{equation}
\begin{equation}
 a_0f=
\left(
\begin{array}{ccc}
 0. & -0.0000234 & -3.011\times 10^{-12} \\
 0.0000234 & 0. & 2.843\times 10^{-12} \\
 3.011\times 10^{-12} & -2.843\times 10^{-12} & 0. \\
\end{array}
\right)
\end{equation}
This reproduces the neutrino observables $\Delta m_{21}^2 = 7.22 \times 10^{-5} \mathrm{eV}^2$, $\Delta m_{31}^2 = 2.449 \times 10^{-3} \mathrm{eV}^2$, $\theta_{12}=33.04^\circ  $, $\theta_{23}=51.89^\circ $ and $\theta_{13}=8.88^\circ$ consistent within $3 \sigma$ with the best fit value from global analysis \cite{Esteban:2020cvm}. We are interested in a scenario wherein $M_{H^+} \approx M_{H} \approx M_{A} \approx 2$ TeV, $M_{h^+} \approx 220$ GeV, and $\sin{\omega} \approx 0.35$. In this context, the NSIs induced by $h^+$ can reach $\epsilon_{\tau \tau}^{\mu \mu (h^+)} \approx 19\%$ 
making it a feasible target for exploration in future muon colliders, as depicted in Fig.~\ref{Fig: Sensitivity UV models}.
While we primarily study charged scalar-mediated NSI in the Zee model, our results on NSI sensitivity are also applicable to other models like the 2HDM \cite{Branco:2011iw} or other radiative neutrino mass models \cite{Babu:2019mfe} featuring charged scalars.

\vspace{0.05 in}
{\textbf {\textit {Conclusions.--}}}
We have analysed the monophoton signal $\mu^+ \mu^- \rightarrow \nu \overline{\nu} \gamma$ to determine the sensitivity of a high-energy muon collider to neutrino NSIs, with an emphasis on muonic forces. Our work ranges from effective field theories to illustrative UV completions. We have shown that a $\sqrt{s} = 3 $ TeV muon collider with integrated luminosity $\mathcal{L} = 1 ~ \mathrm{ab}^{-1}$ has the potential to probe NSIs with strength as big as $|\epsilon_{\tau\tau}^{\mu\mu}| \gtrsim 1.5 \times 10^{-4}$ at $95 \% $ CL.  Intriguingly, this could probe previously unexplored regions of the parameter space of (a) the $L_\mu - L_\tau$ model, (b) Zee's neutrino mass model, and (c) the muon-philic scalar model, which has the potential to explain the discrepancy between observed and predicted muon magnetic moments.

\vspace{0.05in}
{\textbf {\textit {Acknowledgments.--}}}
 We thank K.S. Babu, André de Gouvêa and Guoyuan Huang  for the discussions. 

\appendix
\section*{Supplemental Material}
\subsection{Differential cross sections}
\begin{figure}[h!]
		\centering
		\includegraphics[width=0.35\textwidth]{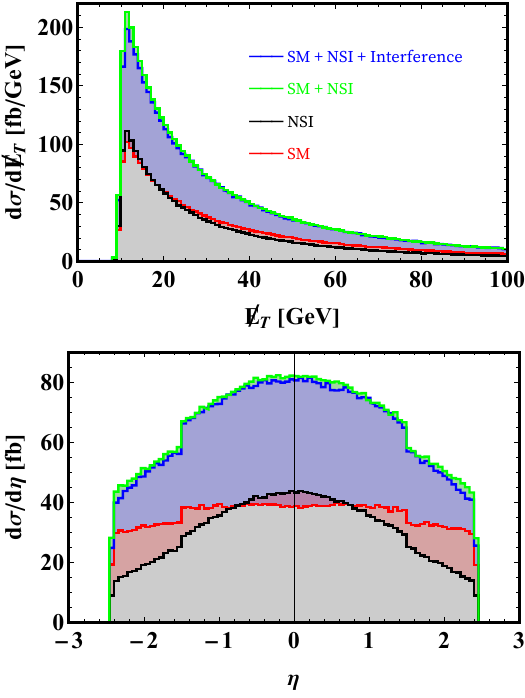}
	\caption{Differential cross sections $d\sigma/d\slashed{E}_T$ and $d\sigma/d\eta$ at a $\sqrt{s} = 3$ TeV muon collider for the process $\mu^+ \mu^- \rightarrow \overline{\nu} \nu \gamma $ for the SM (red), neutrino NSI alone (black) and both contributions including also interference effects (blue). The green line gives the sum of SM and NSI contributions and shows that interference effects are small. The neutrino NSI is induced by the simplified $Z'$ model given in Eq.(\ref{Eq.: Z' simplified model lagrangian}) with $M_{Z'}= 3$ TeV, $\Gamma_{Z'}= 0.3$ TeV   and  couplings $g^L_\mu = g^R_\mu =1 $ and $ g_\nu^{\alpha \beta} = \delta^{\alpha \tau} \delta^{\beta \tau} $.}
	\label{Fig:Interference cross check}
\end{figure}

Fig.~\ref{Fig:Interference cross check} displays the differential cross sections $d\sigma/d\slashed{E}_T$ and $d\sigma/d\eta$ from Monte Carlo simulations of $\mu^+ \mu^- \rightarrow \overline{\nu} \nu \gamma$ process. The red curve represents the Standard Model, the black curve corresponds to NSI per Eq.(\ref{Eq.: Z' simplified model lagrangian}), and the blue curve dictates the total effect of both Standard Model and NSI, with interference considered. To facilitate comparison, the green curve indicates the combined contributions of the Standard Model and NSI, excluding interference effects.

\subsection{Details of the  Zee model}
Here we  briefly summarize the Zee model \cite{Zee:1980ai}.
In the scalar sector, the Zee model consists of  two Higgs doublets $H_{1,2}\sim(1,2,1/2)$ and a charged scalar singlet $\eta^+ \sim (1,1,2)$ with the $SU(3)_C \times SU(2)_L \times U(1)_Y$ charges given in brackets. The charged scalar singlet couples to the weak lepton doublet via 
\begin{equation}
    -\mathcal{L}_Y \supset f_{\alpha \beta} L_\alpha \epsilon L_\beta \eta^+ + h.c.~,
\end{equation}
where $\epsilon$ is the Levi-Civita tensor acting on $SU(2)_L$ indices and $f_{\alpha \beta}$ the Yukawa coupling matrix that is anti-symmetric in the flavor indices $\alpha,~ \beta$. In the Higgs basis \cite{Babu:2018uik}, where only $H_1$ acquires a vacuum expectation value $v$ in its neutral component, the Yukawa couplings of the two Higgs doublets are given by
\begin{equation}
    -\mathcal{L}_Y \supset \tilde{Y}_{\alpha \beta} L_\alpha \Tilde{H}_1  \ell^c_{\beta} \epsilon + Y_{\alpha \beta} L_\alpha \Tilde{H}_2  \ell^c_{\beta } \epsilon + h.c. ~,
\end{equation}
where  $\Tilde{H}_{1,2} = i \sigma_2 H^*_{1,2} $, $\ell^c_{\alpha }$ denotes the left-handed antilepton field and $ \tilde{Y}_{\alpha \beta} $, $Y_{\alpha \beta}$ are $3 \times 3$ Yukawa coupling matrices. Together with a cubic term in the potential, $-V \supset \mu H_1 \epsilon H_2 \eta^- +h.c.$, this leads to the neutrino mass
\begin{equation}
\begin{split}
    M_\nu &= a_0 \left( f m_E Y_l- Y_l^T m_E f\right)~, \\
    a_0 = &\dfrac{\sin 2 \omega}{16 \pi^2} \log\left(\dfrac{M^2_{h^+}}{M^2_{H^+}} \right)~,~ \sin 2\omega = \dfrac{\sqrt{2}v \mu}{M^2_{h^+}-M^2_{H^+}} ~,
\end{split}
\end{equation}
where we work in a basis where $m_E = \mathrm{diag}\left(m_e,~ m_\mu,~ m_\tau \right)$ and $\omega$ denotes the mixing angle  between the two singly charged scalars $\eta^+$ and $H_2^+$ yielding the mass eigenstates $h^+$ and $H^+$, given by:
\begin{equation}
\begin{aligned}
h^{+} & =\cos \omega \eta^{+}+\sin \omega H_2^{+} \\
H^{+} & =-\sin \omega \eta^{+}+\cos \omega H_2^{+}
\end{aligned}
\end{equation}

Besides the two singly charged scalars, the physical scalar spectrum includes two neutral CP-even scalars $h$ and $H$, with the former being identified as the SM Higgs, and a  CP-odd scalar $A$.
Lepton flavor violating (LFV) decays  can be mediated by  the new scalars of the model and therefore impose various constraints on the Yukawa couplings. By considering no more than one large entry in a row of $Y_l$ and $f$, dangerous  LFV decays mediated by the light-charged scalar $h^+$ can be avoided. However, achieving a  viable phenomenology in the neutrino sector necessitates the existence of further non-zero Yukawa couplings, except for those needed to have large NSI.  For the given benchmark point, the sizable $Y_{\tau \tau}$ contribution leads, for example, to a non-vanishing branching ratio of the radiative decay $\tau \rightarrow \mu \gamma$ mediated by the charged scalars \cite{Babu:2019mfe}
\begin{equation}
\begin{split}
    \mathrm{Br} (\tau \rightarrow \mu \gamma)   &\simeq \frac{1}{\Gamma_\tau} \dfrac{\alpha}{4} \dfrac{|Y_{\tau \tau} Y^*_{\tau \mu}|^2}{(16 \pi^2)^2} \dfrac{m_\tau^5}{144}\left(\dfrac{\sin^2 \omega}{M_{h^+}} +  \dfrac{\cos^2 \omega}{M_{H^+}}\right)^2 \\
    &\simeq 3.0 \times 10^{-9}
    \end{split}
\end{equation}
where $\Gamma_\tau $ is the total decay width of tau lepton, and the result is in agreement with the latest PDG limit $\mathrm{BR}(\tau \rightarrow \mu \gamma ) < 4.4 \times 10^{-8}$ \cite{ParticleDataGroup:2022pth}. 
Similarly, considering constraints for all other LFV processes of the type $\ell_\alpha \rightarrow \ell_\beta \gamma $ and $\ell_\alpha \rightarrow \overline{\ell}_\beta \ell_\gamma \ell_{\delta} $ we find our benchmark point in agreement with current experimental bounds.  
Besides, new decay channels can arise in the Zee model, where the muon and tau decay to a charged lepton and two neutrinos. For instance, the process $\mu^- \rightarrow e^- \overline{\nu_\mu} \nu_\tau $ has a branching ratio
\begin{equation}
\begin{split}
    \mathrm{Br}(\mu^- \rightarrow e^- \overline{\nu_\mu} \nu_\tau) &\simeq \frac{1}{\Gamma_\mu} \dfrac{2}{6144  \pi^3} \dfrac{m_\mu^5}{M_{h^+}^4} \left( \sin \omega \cos\omega Y_{\tau \mu} f_{\mu e} \right)^2\\
    &\simeq 1.7 \times 10^{-7}
\end{split}
\end{equation}
For LFV muon decays, the current limit on  $\mu \rightarrow e \nu \nu$ is given by the PDG $\mathrm{Br}(\mu^- \rightarrow e^- \overline{\nu_\mu} \nu_e) < 1.2 \%$ \cite{ParticleDataGroup:2022pth} (See \cite{Kuno:1999jp}, for details of muon anomalous decay) which is not so stringent. Tighter bounds originate from the general fit to muon and tau decay parameters. For the muon, the charged scalars in the Zee model contribute to the operator $\mathcal{L}_{\mathrm{eff}} = 2 \sqrt{2} G_F g_{RR}^S (\overline{\nu}_{\alpha L } e_R) (\overline{\mu}_R \nu_{\beta L})$ (compare to notation in \cite{ParticleDataGroup:2022pth}). For $h^+$ mediated decay we can identify $g_{RR}^S = Y_{\alpha e} Y^*_{\beta \mu} \sin^2{\omega}/(2 \sqrt{2} G_F M_{h^+}^2)$ which is well below the current PDG limit $|g_{RR}^S| < 0.035$ \cite{ParticleDataGroup:2022pth} for our benchmark point, and we checked that similar results hold true for tau decay parameters. For other experimental constraints, we follow Ref.~\cite{Babu:2019mfe}, and we are consistent with the electroweak $T$ parameter, SM Higgs observable, and charge-breaking minima constraints.

\bibliographystyle{utcaps_mod}
\bibliography{reference}
\end{document}